\documentclass{PoS}
\usepackage{mciteplus}
\usepackage{url}

\newcommand{\HS}{{\sc HiggsSignals}}
\newcommand{\HB}{{\sc HiggsBounds}}
\newcommand{\FH}{{\sc FeynHiggs}}
\newcommand{\GeV}{\; \mathrm{GeV}}

\title{Constraining extended Higgs sectors\\ with HiggsSignals}

\ShortTitle{Constraining extended Higgs sectors with HiggsSignals}
\dedicated{We dedicate this paper to Emma St{\aa}l Werlefors, born on 20 July 2013.}

\ReportNo{BONN-TH-2013-20}

\author{Oscar St{\aa}l\footnote{Registered speaker.}\\
        The Oskar Klein Centre\\ Department of Physics, Stockholm University SE-106 91 Stockholm, Sweden\\
        E-mail: \email{oscar.stal@fysik.su.se}}
\author{Tim Stefaniak\footnote{Actual speaker.}\\
        Bethe Center for Theoretical Physics\\
         Physikalisches Institut der Universit\"at Bonn, 
         Nu{\ss}allee 12, D-53115 Bonn, Germany\\
        E-mail: \email{tim@th.physik.uni-bonn.de}}

\abstract{

We introduce the public computer code \HS, which can be used to test the predictions from models with arbitrary Higgs sectors against experimental measurements. Following a brief description of the code, several examples of \HS\ applications are given. We derive constraints on extended Higgs sectors taking into account the measured rates for the observed LHC Higgs signal in all available channels. The results are presented both in a model-independent framework with coupling scale factors that parametrize deviations from the Standard Model, and in a model-dependent example where we analyze a specific benchmark scenario of the minimal supersymmetric standard model.}

\FullConference{The European Physical Society Conference on High Energy Physics - EPS-HEP2013\\
		18-24 July 2013\\
		Stockholm, Sweden}

\begin{document}

\section{Introduction}
Following the discovery of a Higgs-like boson by the LHC experiments ATLAS \cite{ATLASDiscovery} and CMS~\cite{CMSDiscovery}, there has been vivid activity both in the experimental and theory communities to fit the measured properties of this new state---most notably its mass and the production times decay rates---in various models. For a few examples of such analyses, see e.g.~Refs.~\cite{ATLAS-CONF-2013-034,*CMS-PAS-HIG-13-005,Espinosa:2012im,*Giardino:2013bma,*Ellis:2013lra,*Belanger:2013xza,*Lopez-Val:2013yba}.

The public computer code \HS\ \cite{Bechtle:2013xfa} has been developed to provide a coherent framework for comparing theoretical Higgs sector predictions of arbitrary models to this experimental data. Based on model input for the Higgs sector (supplied by the user) and published experimental results from ATLAS, CMS, and the Tevatron experiments, \HS\ tests the compatibility between data and theory for any model that fulfills a few basic assumptions (for example that the narrow width approximation holds).

This paper serves as a brief introduction to the \HS\ program and demonstrates some first applications. For a full account of the implemented physics, statistical procedure, and practical user instructions, we refer to the manual \cite{Bechtle:2013xfa}. The code can be obtained from the webpage:
\begin{itemize}
\centering
\vspace{-0.5em}
\item[] \url{http://higgsbounds.hepforge.org}
\end{itemize}

\section{Basics of HiggsSignals}
\HS\ has been developed to work seamlessly together with its sister program, \HB\ \cite{Bechtle:2008jh,*Bechtle:2011sb,*Bechtle:2013gu}. \HB\ compares model predictions to exclusion limits from direct Higgs searches at LEP, the Tevatron and the LHC. With the discovery of a Higgs-like state, complementary, and often more precise, constraints can be obtained by comparing to measurements of the observed signal, rather than to exclusion limits. \HS\ therefore provides an essential extension to this code. The basic model input used by \HS\ is the same as for \HB: the number of (neutral) Higgs bosons, their masses, production cross sections, total widths, and branching ratios into various final states. Several options are available for specifying the cross sections, which can be given as full hadronic cross sections, as partonic cross sections, or in terms of effective couplings. In the latter case the model cross sections are approximated by a simple rescaling of the corresponding SM quantities. One extension of the input made for \HS\ is that theoretical uncertainties can be specified, both on the Higgs masses as well as the different Higgs production processes and decays. This is important to get realistic results for models where these differ from the SM case. For a user of \HB, the input format is familiar, and it should be straightforward to get started with \HS. Several example programs are also provided.

\begin{figure}
\centering
\vspace{-1em}
\includegraphics[width=0.38\columnwidth]{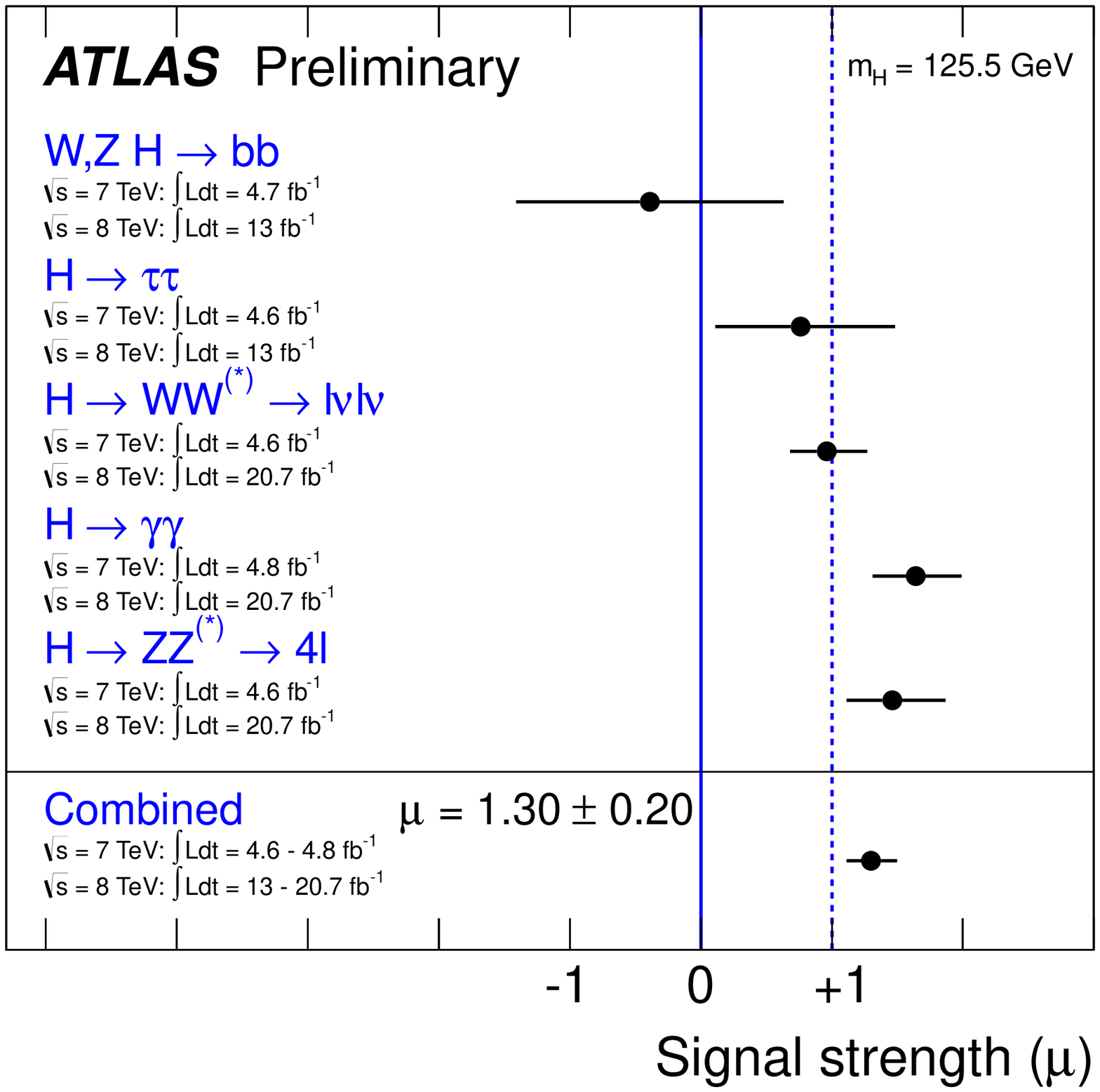}
\includegraphics[width=0.39\columnwidth]{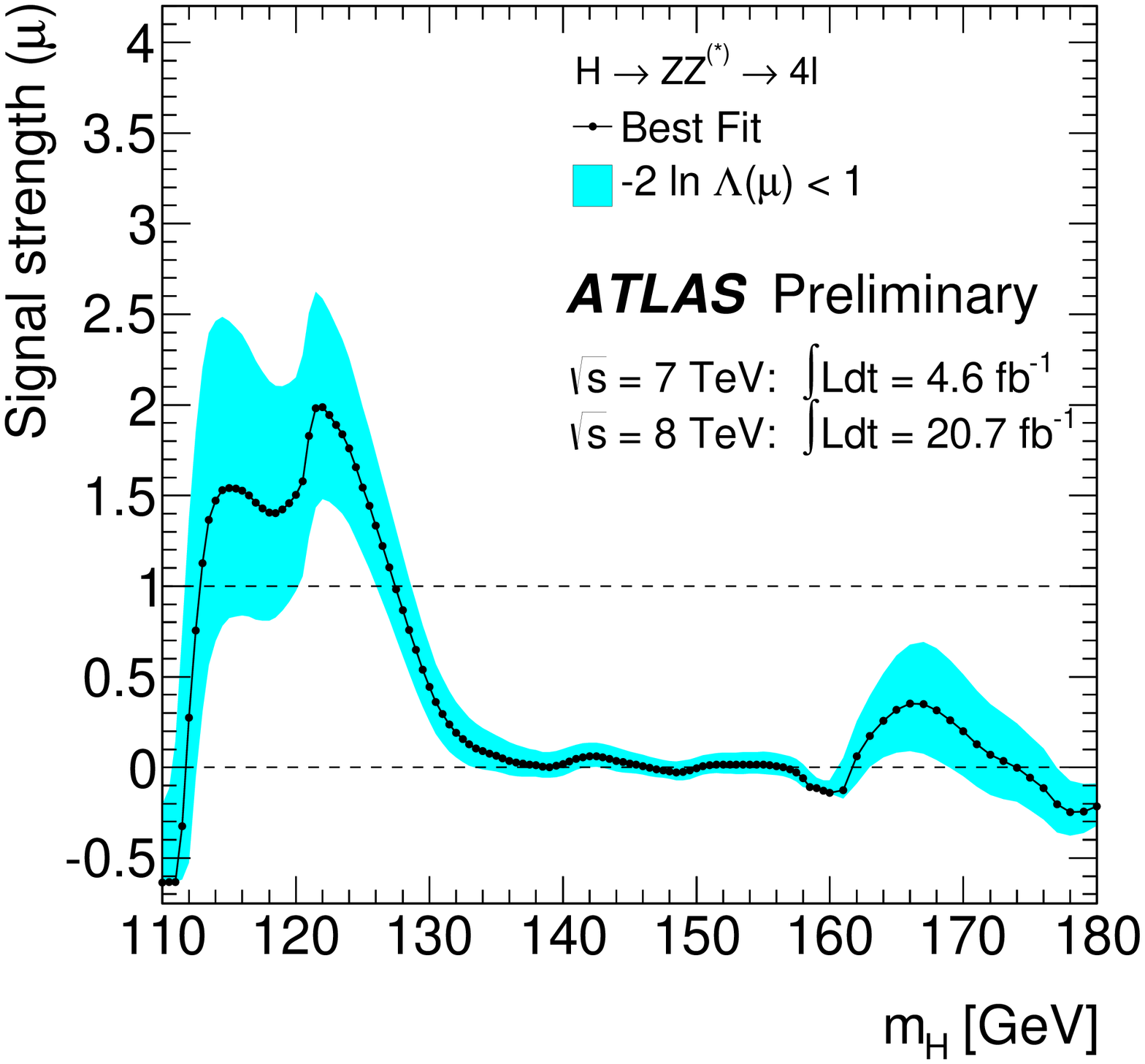}
\vspace{-0.5em}
\caption{Example of the experimental results used in \HS: ATLAS signal strength modifier in different channels at fixed Higgs mass, $m_H$ \cite{ATLAS-CONF-2013-034} (left), and in $H\to ZZ^{(*)}\to 4\ell$ as a function of $m_H$ \cite{ATLAS-CONF-2013-013} (right).}
\label{fig:muplots}
\end{figure}

Together with the measured Higgs masses in different channels, the basic experimental quantity used in \HS\ is the \emph{signal strength modifier}, $\mu_{xx}$,  which represents a measurement of the signal rate in a particular final state, $xx$, 
 normalized to the SM prediction. The measurement can either be fully inclusive, or targeting a specific Higgs production mechanism.
 These measurements are performed both as a function of the Higgs mass, and for specific (interesting) values of $M_H$ where a signal is observed, see Fig.~\ref{fig:muplots}. An obvious example of the latter is the observed LHC signal around $M_H\sim 125\GeV$, but it should be stressed that \HS\ is completely general and can work with data sets containing any number of measured ``signals'' in different experiments, including toy data. A number of different data sets are provided with the code. The default is called \emph{latestresults} and contains all the latest public results from ATLAS, CMS, and the Tevatron experiments. New data sets can be specified by the user in a simple text format. 

When using \HS, the experimental measurements of the signal strengths are statistically compared to the model predictions defined by
$\mu_{xx} = \sum_i \omega_i \mu_{xx}^i$. Here $\omega_i$ is the relative contribution to the signal from a particular Higgs production process $i$. The weights, which include the experimental efficiencies, are inferred from the published information about the analyses (where available).  The individual signal strengths, $\mu^i_{xx}$, for each production mode are calculated from the model input as
\begin{equation}
\mu^i_{xx}=\frac{\left[\sigma_i(pp\to H)\times \mathrm{BR}(H\to xx)\right]_{\mathrm{model}}}{\left[\sigma_i(pp\to H)\times \mathrm{BR}(H\to xx)\right]_{\mathrm{SM}}}.
\end{equation}
To compare the theory predictions to the experimental measurements, \HS\ contains two distinct statistical methods which use different types of data. The user can specify to run either of these methods or to combine them.\newline
 \underline{The peak-centered $\chi^2$ method} uses data at fixed values of the Higgs mass, which are thought to correspond to signals (or ``peaks''), cf.~Fig.~\ref{fig:muplots} (left). These peak masses may however vary between different experimental analyses. The test proceeds by \emph{assigning}, for each peak, Higgs bosons that have a mass in agreement with the peak position within the experimental resolution. Following the assignment, a $\chi^2$ measure is evaluated by comparing the signal strength measurement for the peak to the predicted $\mu$ value. If a mass measurement is available\footnote{This is so far the case for the high-resolution channels $H\to 4\ell$ and $H\to \gamma\gamma$.} the corresponding $\chi^2$ contribution from the comparison between the predicted and observed Higgs masses is added. In a situation where multiple Higgs bosons contribute to one signal, and also in the (somewhat unlikely) case where the same analysis displays multiple signals, an optimal assignment of the Higgs bosons to the signals is achieved by minimizing the overall $\chi^2$ function. The signal strengths of multiple Higgs bosons are then added incoherently, which assumes interference is negligible.\newline
\underline{The mass-centered $\chi^2$ method} on the other hand uses the $\mu(m_H)$ measurements, cf.~Fig.~\ref{fig:muplots} (right), and performs a $\chi^2$ evaluation of the model rate predictions against the data directly at the predicted value of the Higgs mass(es). This approach is therefore complementary to the peak-centered $\chi^2$ method, since it does not measure compatibility of the model with particular designated signals. Also here, \HS\ combines (incoherently) the signal strengths of multiple Higgs bosons contributing in experimentally indistinguishable mass regions, see \cite{Bechtle:2013xfa} for details. The mass-centered $\chi^2$ method is limited in applicability by the much fewer experimental results of this type that are available. 

\vspace{-0.5em}
\section{Example applications}
\vspace{-0.5em}
To demonstrate the use of \HS, we now show results from different fits produced with the peak-centered $\chi^2$ method. All the available data for the Higgs signal at $\sim 125$~GeV from ATLAS, CMS, and the Tevatron presented until the Moriond 2013 conference has been included. 

\begin{figure}
\vspace{-1em}
\centering
\includegraphics[width=0.45\columnwidth]{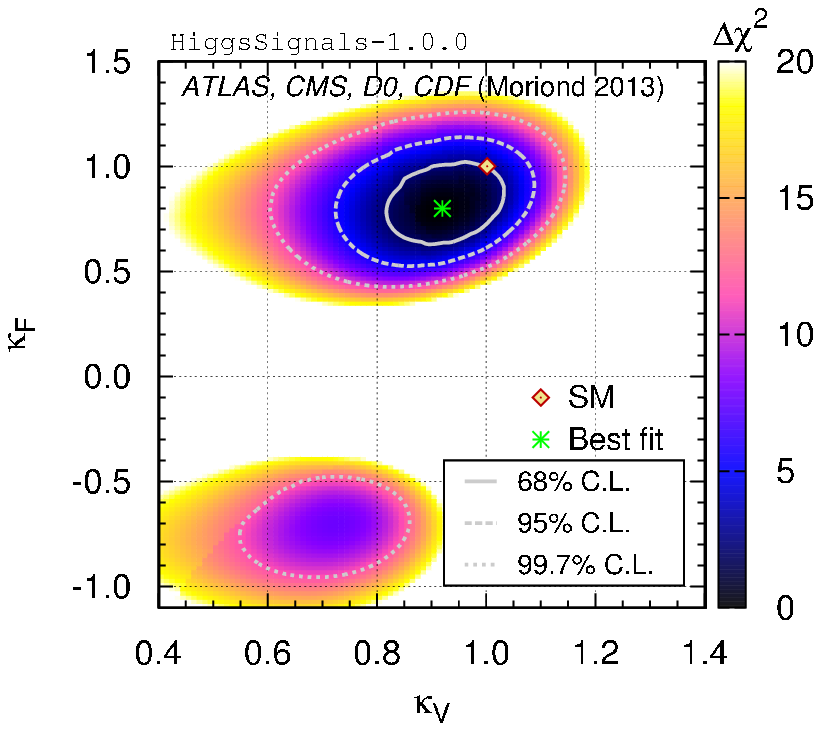}\, \, \, \,
\includegraphics[width=0.43\columnwidth]{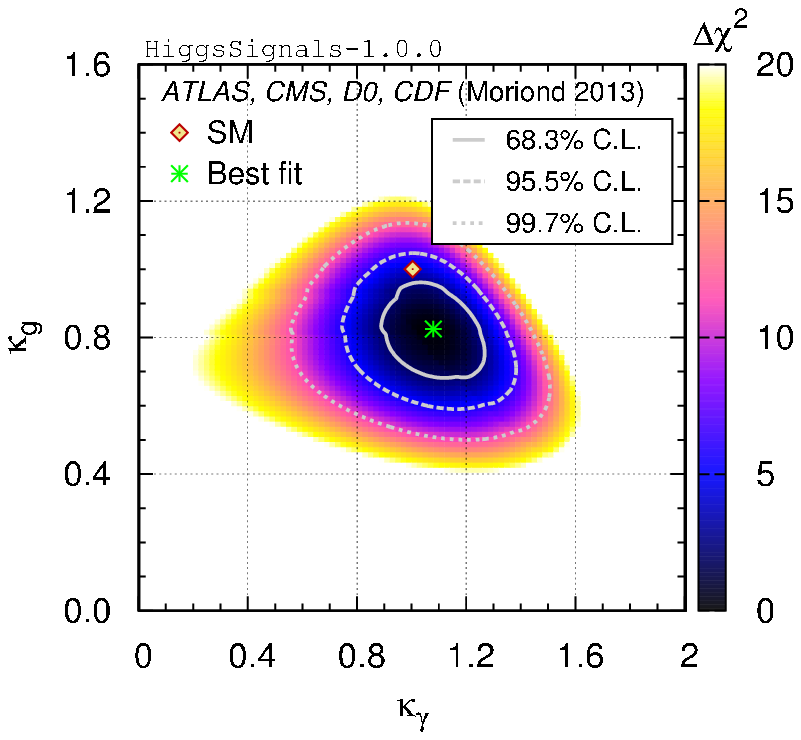}
\caption{Fit results using \HS\ with all available data after Moriond 2013 for the SM modified with universal scale factors for Higgs couplings to fermions, $\kappa_F$, and vector bosons, $\kappa_V$ (left); gluons, $\kappa_g$, and photons, $\kappa_\gamma$ (right). The colours indicate levels of $\Delta\chi^2$ from the respective best fit points (green stars).}
\label{fig:app1}
\end{figure}

As a first application, we consider a model with a single Higgs boson where scale factors are allowed to modify the Higgs couplings. These scale factors, denoted by $\kappa_i$,  are defined in Ref.~\cite{LHCHiggsCrossSectionWorkingGroup:2012nn, *Heinemeyer:2013tqa} to parametrize deviations from the SM. The normalization is such that the best available SM theory prediction is recovered when $\forall i:\kappa_i=1 $. 
The results are presented for two different fits. In the first case, a universal scaling is allowed for the Higgs couplings to fermions, $\kappa_F$, and to the weak gauge bosons, $\kappa_V$ ($V=W,Z$). The loop-induced couplings to gluons and photons are then calculated from the indvidual contributions. The results are shown in Fig.~\ref{fig:app1} (left). We obtain the best fit point $(\kappa_V, \kappa_F)=(0.92,0.80)$ with $\chi^2_{\mathrm{min}}/\mathrm{ndf}=31.0/43$. The combined data shows no significant deviation from the SM, and the solution with $\kappa_F < 0$ is disfavored by more than $2\,\sigma$.
In the second case, the loop-induced Higgs couplings to gluons and photons are allowed to be rescaled by $\kappa_g$ and $\kappa_\gamma$, respectively. This can be understood as an effective description of new physics contributions entering only in the loop diagrams generating these couplings. The results from this fit are shown in Fig.~\ref{fig:app1} (right). The best fit is found at $(\kappa_\gamma, \kappa_g)=(1.08,0.83)$ with $\chi^2_{\mathrm{min}}/\mathrm{ndf}=29.7/43$. Again, there is no significant deviation from the SM.

\begin{figure}
\vspace{-0.5em}
\centering
\includegraphics[width=0.4\columnwidth]{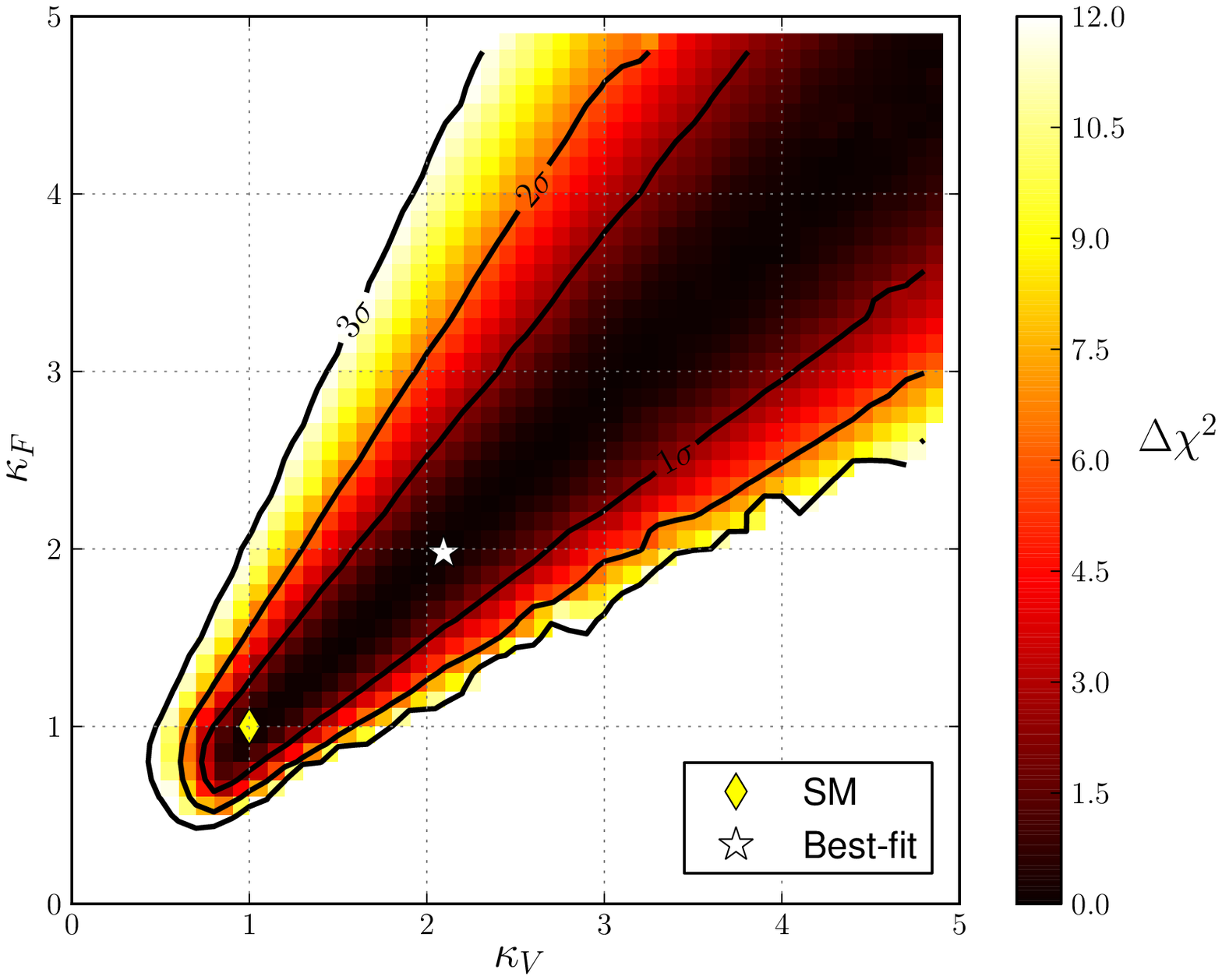}
\includegraphics[width=0.4\columnwidth]{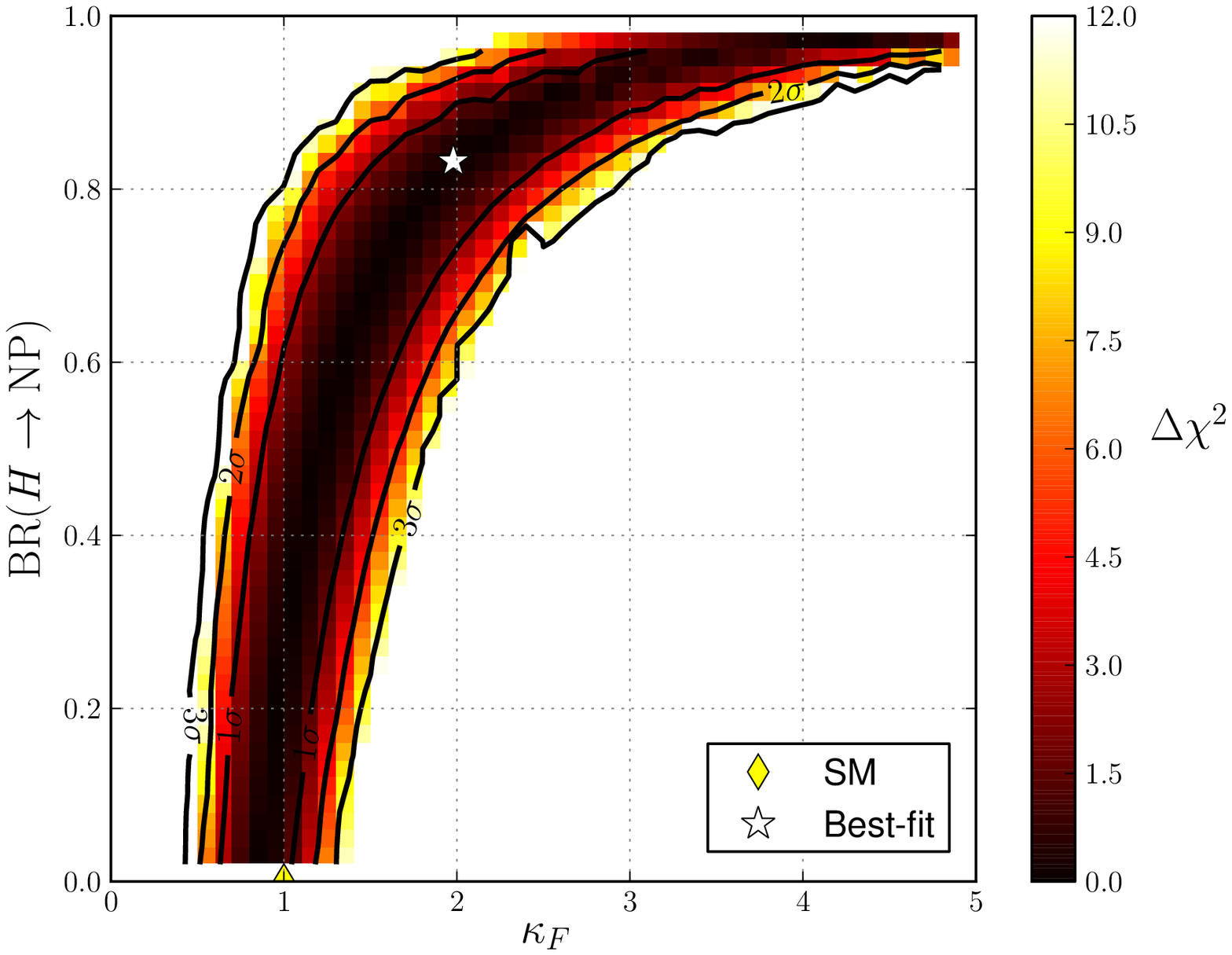}
\includegraphics[width=0.4\columnwidth]{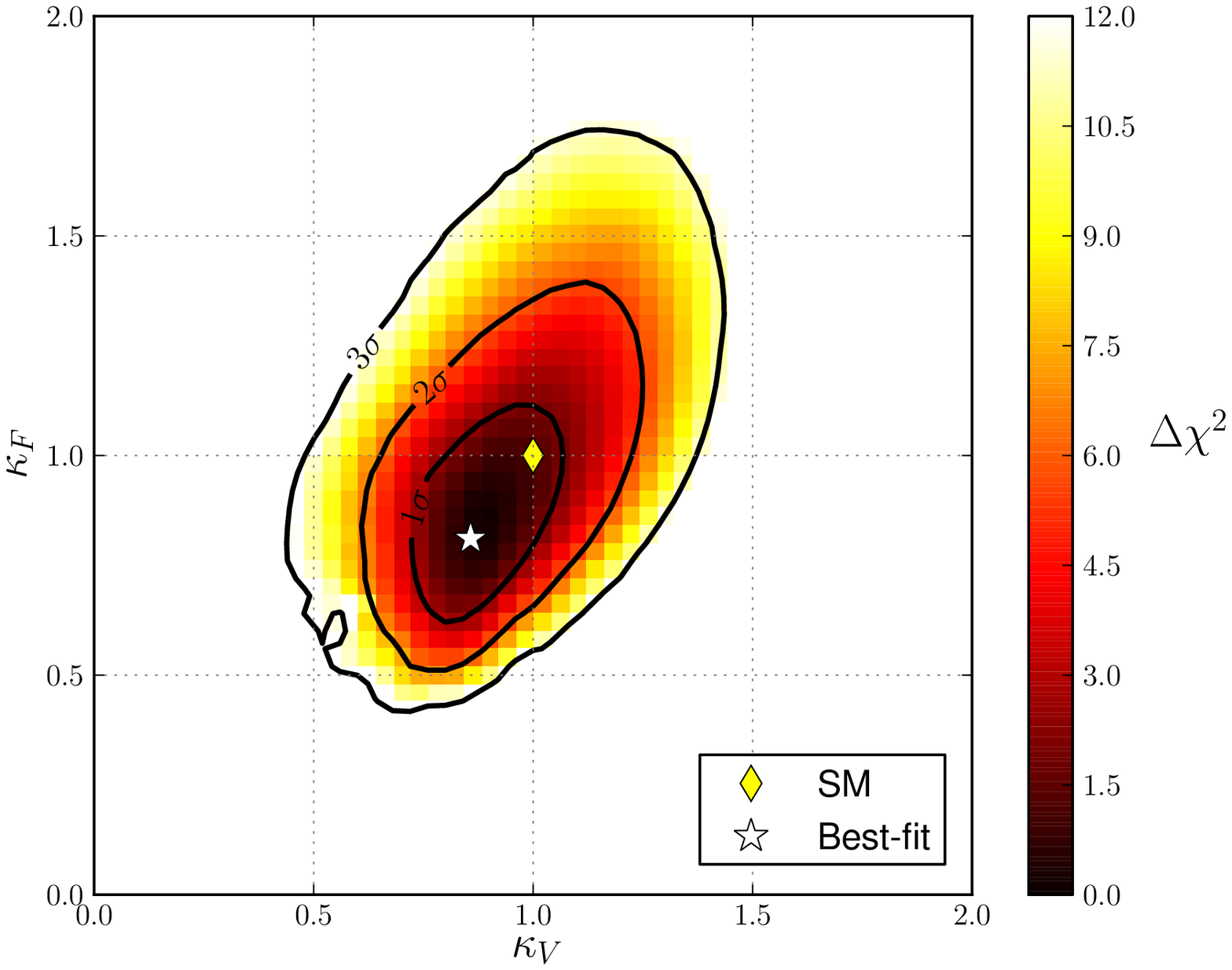}
\includegraphics[width=0.4\columnwidth]{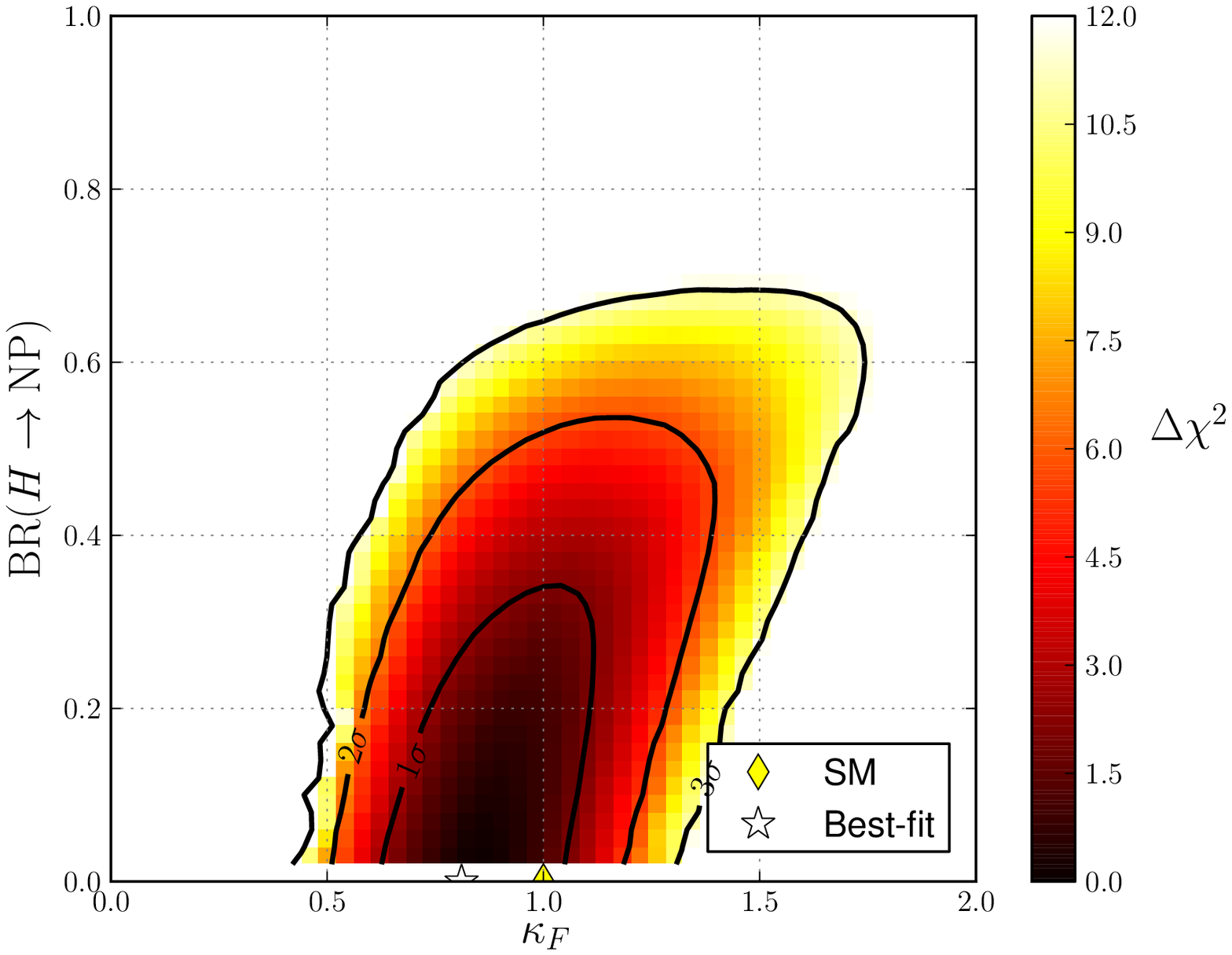}
\caption{\HS\ fit to universal coupling scale factors for fermions and vector bosons with the possibility of undetected Higgs decay modes. The fit performed without restrictions on the Higgs total width (upper), and taking into account constraints on $\mathrm{BR}(H\to \mathrm{invisible})$ from ATLAS \cite{ATLASinv} (lower). The colours indicate levels of $\Delta\chi^2$ from the respective best fit points (white stars).}
\label{fig:app2}
\vspace{-1em}
\end{figure}

In a second \HS\ application, we start from the simple model with two coupling scale factors, $\kappa_V$ and $\kappa_F$, but allow in addition for a new, undetected, decay mode of the Higgs boson. This is described by an additional branching ratio $\mathrm{BR}(H\to \mathrm{NP})$, and the resulting total Higgs width can be written as $\Gamma_H=\Gamma_H^{\mathrm{SM}}/(1-\mathrm{BR}(H\to \mathrm{NP}))$. Including this decay affects the fit results quite dramatically, as can be seen in Fig.~\ref{fig:app2} (upper row). The presence of the new unconstrained decay mode causes a degeneracy between an increased Higgs production in all modes (for $\kappa_F \gg 1$, $\kappa_V\gg 1$), and a simultaneous increase of this new undetectable decay mode, such that the overall rates in the observed channels are close to their SM values. This is a manifestation of the known fact that the total decay width $\Gamma_H$ is not directly accessible to measurement at the LHC.
With additional assumptions, this degeneracy can be broken. One example is when the new decay mode leads to missing transverse energy, also known as an \emph{invisible} decay, e.g.~when the Higgs boson decays to a pair of stable dark matter particles. In this case, there is a direct limit on $\mathrm{BR}(H\to \mathrm{invisible})$ from the Higgsstrahlung process \cite{ATLASinv} that can be applied. Using the likelihood information in \cite{ATLASinv} as an additional contribution to the fit, we obtain the result shown in the lower row in Fig.~\ref{fig:app2}. Here we can again observe a more limited allowed region in $(\kappa_V, \kappa_F)$, which is very similar to the previous result in Fig.~\ref{fig:app1}. Fig.~\ref{fig:app2} (right) illustrates that the best fit point favors $\mathrm{BR}(H\to \mathrm{NP})=0$ once this constraint is included, but also that a large invisible decay fraction $\mathrm{BR}(H\to \mathrm{invisible})\simeq 50\%$ is still allowed at the $2\,\sigma$ level.

\begin{figure}
\vspace{-1em}
\centering
\includegraphics[width=0.6\columnwidth]{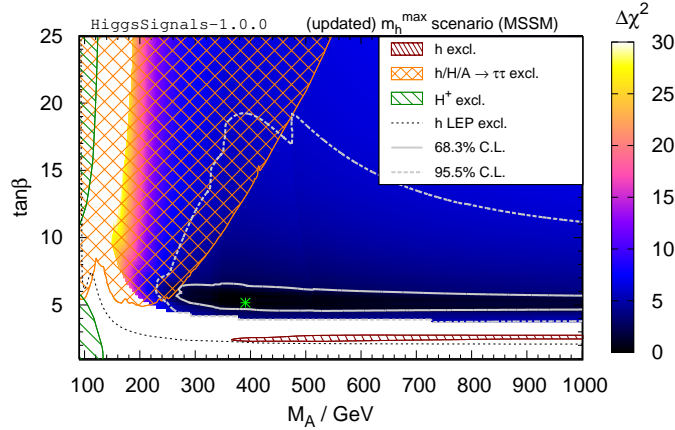}
\vspace{-0.5em}
\caption{Example of \HB/\HS\ application to the updated $m_h^{\mathrm{max}}$-scenario of the MSSM. The hatched regions are excluded by different direct search limits (as indicated by the legend). The colours indicate levels of $\Delta \chi^2$ from the best fit point (green star), and white contours are shown for the $1\,\sigma$ (solid) and $2\,\sigma$ (dashed) regions.}
\label{fig:mssm}
\vspace{-1em}
\end{figure}

As a final example we consider the Minimal Supersymmetric Standard Model (MSSM) which has two Higgs doublets. In the CP-conserving case this leads to two CP-even Higgs bosons, $h$ and $H$ (with $m_h<m_H$), one CP-odd Higgs boson, $A$, and a pair of charged Higgs bosons, $H^\pm$. Two parameters are sufficient to specify the Higgs sector at tree-level: the mass of the CP-odd Higgs boson, $m_A$, and the ratio of vacuum expectation values of the two doublets, $\tan\beta$. To calculate the important radiative corrections to the Higgs masses and mixing, and to make phenomenological predictions, we fix the remaining soft SUSY-breaking parameters at the electroweak scale using the (updated) $m_h^{\mathrm{max}}$ benchmark scenario \cite{Carena:2013qia}. The theory predictions are evaluated with \FH\ \cite{Heinemeyer:1998yj,*Hahn:2009zz} and a theory uncertainty of $2$~GeV is assigned to $m_h$. The results from a combined analysis of the $m_h^{\mathrm{max}}$ scenario with \HB/\HS\ are shown in Fig.~\ref{fig:mssm}.
In this figure, LHC exclusion limits (evaluated using \HB) are displayed as hatched regions, the colours indicating exclusion at $95\%$~CL from different searches. For example, the largest excluded region in this scenario comes from the CMS combined MSSM search for $h/H/A\to \tau\tau$ \cite{CMSHig12050}. Negative results from Higgs searches at LEP \cite{Schael:2006cr} are included as a contribution to the overall $\chi^2$, a new feature of \HB-4, which leads to the extended area with high $\chi^2$ at low $\tan\beta$, where $m_h<114\GeV$ (indicated by a dotted gray line). 

The overall best fit point is found for $m_A=390\GeV$, $\tan\beta=5.3$ and has $\chi^2/\mathrm{ndf} = 33.0/48$. Inside the $2\,\sigma$ favoured region, it is always the lightest MSSM Higgs boson, $h$, which has the right mass ($\sim 125\GeV$ within uncertainties) to be assigned as the LHC signal. In the MSSM decoupling limit ($m_A\gg m_Z$), the couplings of the lightest Higgs boson approach those in the SM. Since the SM is in good agreement with measurements, we expect this limit to give a good fit if also the predicted Higgs mass is close to the measured value. The resulting preferred value is $m_A>260\GeV$, and $\tan\beta$ in the decoupling limit lies in the range $4 < \tan\beta < 10$ (at $95\%$ CL). Whereas a lower limit on $m_A$ derived in the $m_h^{\mathrm{max}}$ scenario is fairly robust \cite{Heinemeyer:2011aa}, the preferred range for $\tan\beta$ is limited to the precise benchmark scenario under discussion. More complete analyses of the low-energy MSSM \cite{Bechtle:2012jw} show that a good fit to the Higgs mass and rates can be obtained over the whole allowed $(m_A,\tan\beta)$ plane by varying the soft-breaking parameters. A first application of \HS\ to constrained supersymmetric models has also been presented at this conference~\cite{Fittino}.

\vspace{-1em}
\section{Conclusion}
\vspace{-0.5em}
\HB\ is an established and convenient tool to apply exclusion limits from direct Higgs searches to arbitrary models. As discussed here, its sister code \HS\ has now been published to take into account LHC/Tevatron \emph{measurements}. \HS\ evaluates a $\chi^2$ measure to determine quantitatively the compatibility between data and theory. It has been validated
against official coupling fits, and we have presented first applications to a single Higgs scenario with different coupling scale factors, including new decay modes, and to the MSSM in the context of a particular benchmark scenario.

Our general strategy for the future development of \HS\ is to take all public information into account, and to keep the code up-to-date with the latest experimental results. It will therefore continue to offer phenomenologists a flexible way to constrain models with extended Higgs sectors from measurements at the Tevatron, LHC and future facilities. User feedback is welcome!
%

\emph{
We thank the other members of the \HS\ team: P.~Bechtle, S.~Heinemeyer and G.~Weiglein, for collaboration on this project.}

\vspace{-0.5em}
\bibliographystyle{JHEP}

\bibliography{HiggsSignals}

\end{document}